\documentclass[doublecol]{epl2} 
\begin{document}

\title{Presence of temporal dynamical instabilities in topological insulator lasers}
\shorttitle{Dynamical instabilities in  ... } %Insert here a short version of the title if it exceeds 70 characters

\author{S. Longhi \inst{1,2}, Y. Kominis \inst{3} \and V. Kovanis \inst{4}}
\shortauthor{S. Longhi}

\institute{                    
  \inst{1}  Dipartimento di Fisica, Politecnico di Milano, Piazza L. da Vinci 32, I-20133 Milano, Italy\\
  \inst{2}  Istituto di Fotonica e Nanotecnlogie del Consiglio Nazionale delle Ricerche, sezione di Milano, Piazza L. da Vinci 32, I-20133 Milano, Italy\\
  \inst{3} School of Applied Mathematical and Physical Science, National Technical University of Athens, Athens 15780, Greece\\
  \inst{4} Department of Physics, School of Science and Technology, Nazarbayev University, Astana 010000, Republic of Kazakhstan\\
   }
 {Topological phases}

% 42.55.Ah 	General laser theory
% 42.55.Px 	Semiconductor lasers; laser diodes
% 42.60.Mi 	Dynamical laser instabilities; noisy laser behavior
% Topological phases (quantum mechanics), 03.65.Vf

\abstract{Topological insulator lasers are a newly introduced kind of lasers in which light snakes around a cavity without scattering. Like for an electron current  in a topological insulator material, a topologically protected lasing mode travels along the cavity edge, steering neatly around corners and imperfections without scattering or leaking out. In a recent experiment, topological insulator lasers have been demonstrated using  a square lattice of coupled semiconductor microring resonators with a synthetic magnetic field. However, laser arrays with slow population dynamics are likely to show dynamical instabilities in a wide range of parameter space corresponding to realistic experimental conditions, thus preventing stable laser operation. 
 While topological insulator lasers provide an interesting mean for  combating disorder and help collective oscillation of lasers at the edge of the lattice, it is not clear whether chiral edge states are immune to dynamical instabilities. In this work we consider a realistic model of semiconductor class-B topological insulator laser and show that chiral edge states are not immune to dynamical instabilities.} 
\maketitle

\section{Introduction}
Topological insulators are materials that do not carry electrical currents in the bulk, but do conduct through edge states \cite{r0}. Inspired by similar phenomena in photonics \cite{r0bis}, the idea of topological insulator lasers have been introduced and demonstrated in a series of recent papers \cite{r1,r1bis,r2,r3,r4,r5,r5bis}. Like for electron currents in topological insulators, in such a kind of lasers light can travel unidirectionally along the edge of the optical cavity,  steering neatly around corners and imperfections without scattering or leaking out. In a recent experiment \cite{r5}, a topological insulator laser has been realized, which is based on a square lattice of coupled semiconductor microring resonators with a synthetic magnetic field realized by anti resonant ring links \cite{r6,r7}. The experiment nicely showed that in the topological insulator laser the synthetic magnetic field enables lasing in a chiral edge mode, outperforming the same laser array without the synthetic magnetic field (i.e. in a topological trivial phase) in terms of efficiency, coherence and robustness against disorder. Such a result seem rather promising toward the realization of miniaturized high power and stable laser arrays. However, it is known that, even in the absence of disorder or imperfections, stable laser emission in solid-state or semiconductor laser arrays, corresponding to phase locking, is severely restricted by the onset of dynamical instabilities \cite{r8,r9,r10,r11,r12}, and that to force stable laser emission  in a supermode of the array selection methods, such as those based on non-local laser coupling in Talbot cavities \cite{r13,r14,r15}, are required. The theoretical model of the topological insulator laser, presented in the companion paper \cite{r4}, predicts stable laser emission in a chiral edge supermode when gain saturation is instantaneous (class-A laser). However, in most cases semiconductor lasers show a slow carrier dynamics as compared to the photon lifetime, and thus they belong to class-B lasers \cite{r16}. In class-B laser arrays, dynamical instabilities are very common even when only two lasers are coupled \cite{r8,r11}, with more complex temporal behavior for a larger number of coupled lasers \cite{r10,r12}. Insight can be gained from analytical, numerical and experimental work on optically coupled diode lasers, where we rudimentary expect optical coupling may prevent phase locking and the continuous-wave (cw) operation to be interrupted with stable limit cycles born out of Hopf bifurcations, as well as period doublings that end up in regions of strange chaotic attractors, as key parameters are changed \cite{r8,r10,r16bis,r16tris}. Therefore, a natural question arises: are chiral edge states in a topological semiconductor laser immune to dynamical instabilities as well?\\  
 In this Letter we consider a more realistic model of semiconductor class-B topological insulator laser array and show that, for typical parameter values that apply to experimental conditions, chiral edge states are likely to undergo dynamical instabilities.\\  

\section{Topological insulator laser: model} 
\subsection{Chiral edge states in coupled-microring lattices with synthetic gauge field: Hermitian model}
\begin{figure}
\onefigure[width=7.8cm]{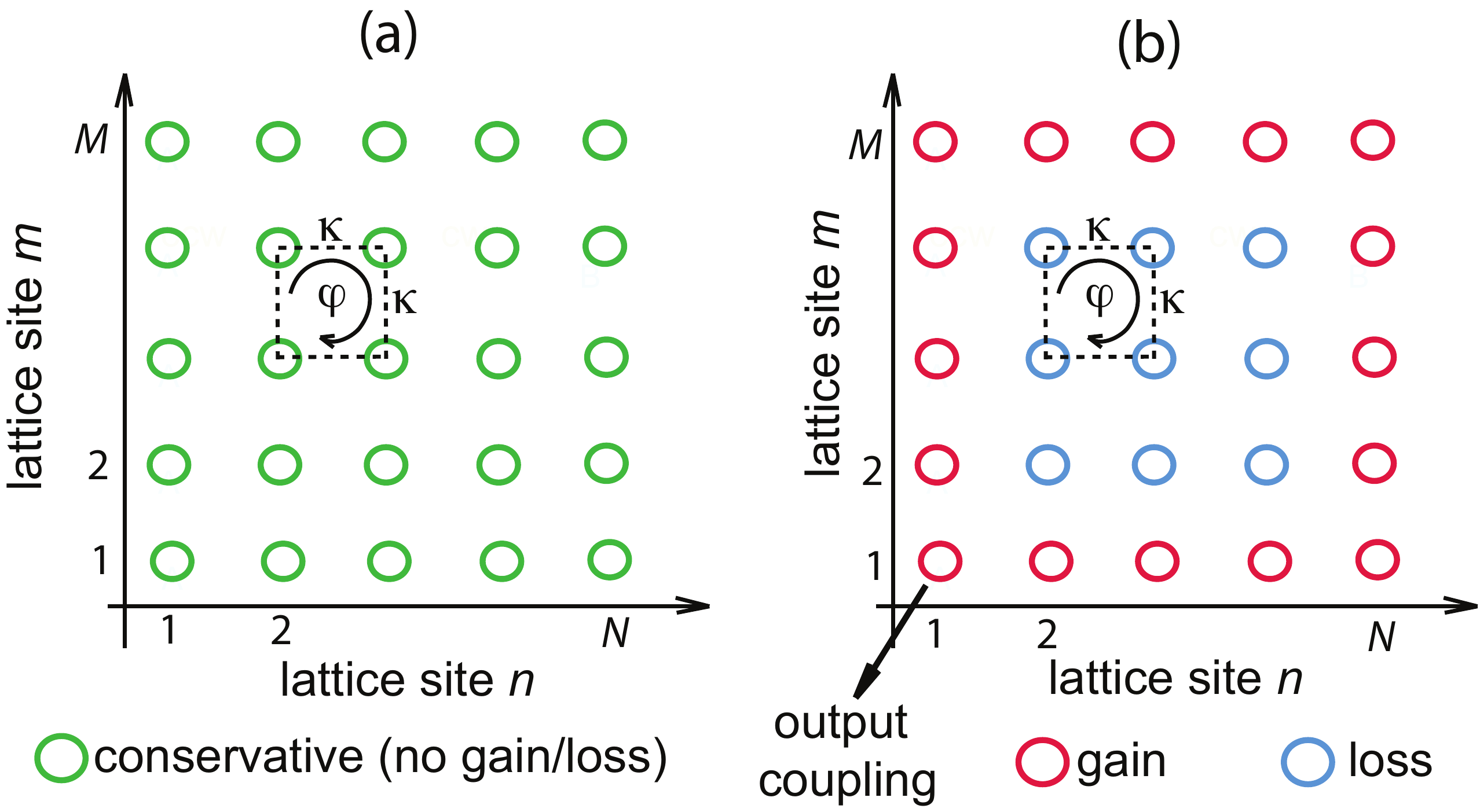}
\caption{(a) Schematic of a $N \times M$ square lattice of evanescently-coupled microring resonators with a synthetic magnetic flux $\varphi$ realized by antiresonant link rings (not shown in the figure) \cite{r7}. We typically assume $M=N$. (b) Schematic of the topological insulator laser \cite{r4,r5}. Optical gain is provided in microrings along the perimeter $\mathcal{P}$ of the lattice, while inner microrings are lossy (to facilitate oscillation in an edge chiral supermode). Light is extracted from the edge microring at site $(1,1)$. }
\end{figure}
The topological insulator laser introduced in \cite{r4} and experimentally realized in \cite{r5} comprises a square lattice of $N \times M$ coupled microrings with a synthetic gauge field $\varphi$ realized by anti resonant link rings; see Fig.1(a). We assume single longitudinal mode oscillation in each microring in a given traveling wave mode; unidirectional oscillation in each microring can be obtained by the method used in  \cite{r5}. Indicating by $c_{n,m}(t)$ the normalized electric field amplitude of the mode oscillating in the $(n,m)$ microring of the lattice and $\kappa$ the effective coupling rate, coupled mode equations in the Hermitian limit, i.e. neglecting gain and loss in the system, read \cite{r7}
\begin{eqnarray}
i \frac{dc_{n,m}}{dt}  & =  & \kappa  \left\{ c_{n+1,m} +c_{n-1,m} \right\} \\
&  + & \kappa \left\{ c_{n,m+1} \exp(i n \varphi) +c_{n,m-1} \exp(-i n \varphi)  \right\} \nonumber
\end{eqnarray}
($n=1,2,...,N$, $m=1,2,...,M$), with open boundary conditions $c_{0,m}=c_{N+1,m}=c_{n,0}=c_{n,M+1}=0$ [Fig.1(a)]. We can write Eq.(1) in the compact form
\begin{equation}
 i \frac{d \mathbf{c}}{dt}=\mathcal {H}^{(Herm)} \mathbf{c},
 \end{equation}
where $\mathbf{c}=(c_{n,m})$ and $\mathcal{H}^{(Herm)}$ is the Hermitian Hamiltonian defined in Eq.(1). For an infinitely extended lattice, the energy spectrum of $\mathcal{H}^{(Herm)}$ depends on  the synthetic gauge phase $\varphi$, i.e. magnetic flux in each plaquette, and realizes the famous Hofstadter butterfly spectrum \cite{r17}. In a finite lattice, chiral edge states, traveling unidirectionally along the perimeter of the lattice, arise in the gapped region of the spectrum. We set the topological phase  $\varphi$ equal to $\varphi= \pi/2$, so that there are four bands with two topological gaps. Figures 2(a) and 2(b) show the computed band diagrams for an infinite stripe ($N=100 \times M=\infty$) and for a square lattice ($N \times N=50 \times 50$), respectively. Clearly, edge modes are found in both cases. In the stripe geometry ($M= \infty$), they can be derived from analysis of Harper equation (see caption of Fig.2). For the finite square lattice ($M=N$) they can be detected by calculating  the participation ratio 
$PR= \left( \sum_{n,m} |c_{n,m}|^2 \right)^2  / \sum_{n,m}|c_{n,m}|^4$ (a larger value of $PR$ corresponds to a more delocalized mode). In the stripe geometry [Fig.2(a)],  each topological gap supports circulation of edge modes in both directions, localized at either edges $n=1$ and $n=N$ of the stripe; the circulation direction is reversed in the two topological gaps. In the finite square lattice the edge states are delocalized all along the perimeter of the square and their energies fall inside the gaps [Fig.2(b)]. \\
 
 \begin{figure}
\includegraphics[width=8cm]{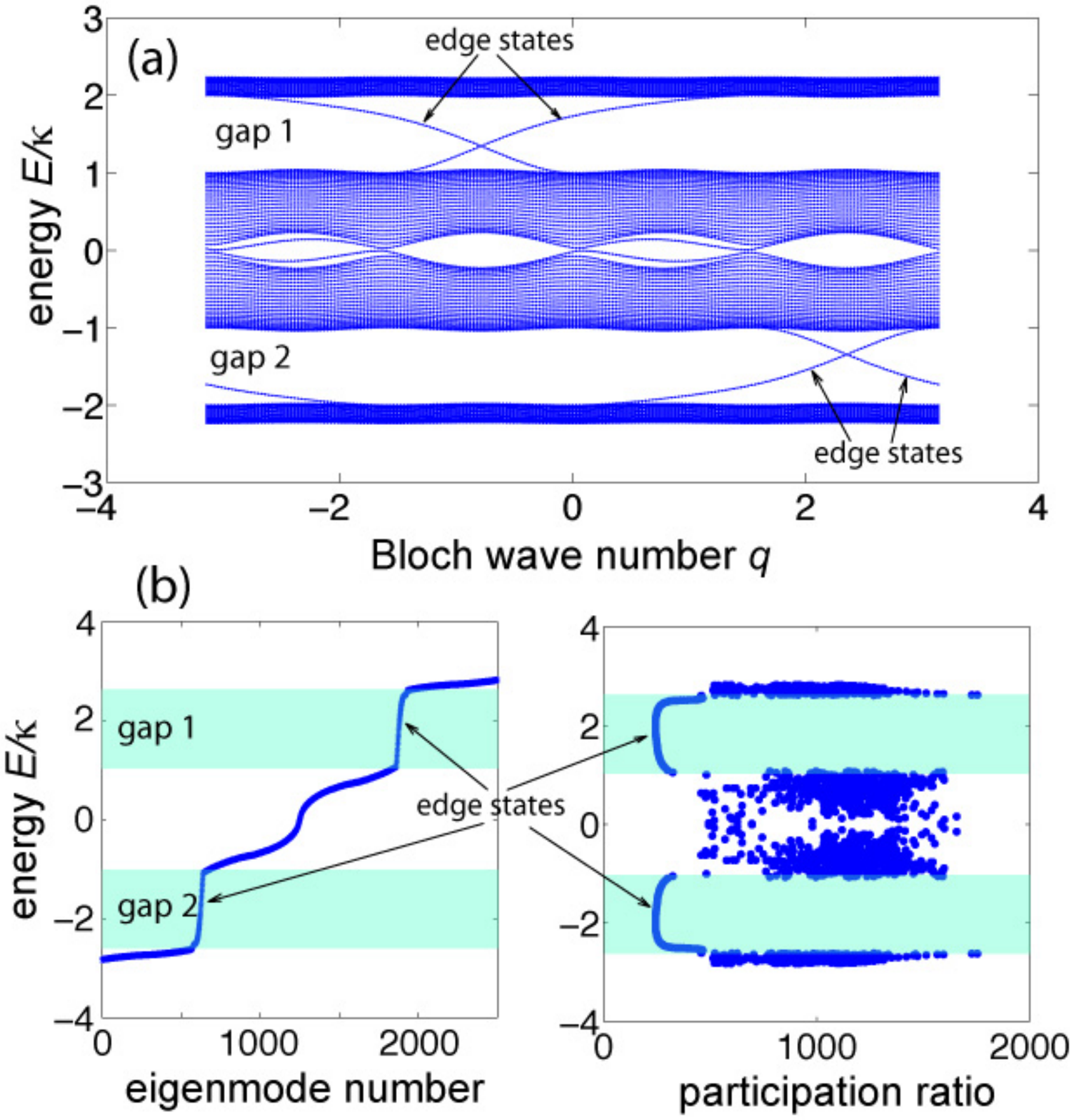}
\caption{ (a) Energy band diagram of a square lattice in a stripe $N \times M=100 \times \infty$ with magnetic flux $\varphi= \pi/2$. Bloch wave number $q$ and energy $E$ are defined by $c_{n,m}=A_n \exp[iqm-i E(q) t]$, where $E=E(q)$ and $A_n$ are the eigenvalues and corresponding eigenvectors of the Harper equation $\kappa(A_{n+1}+A_{n-1})+ 2 \kappa \cos (q+ n \varphi) A_n=E A_n$.  Edge states, with dispersion curves in the two topological gaps, are localized either at the left or right edges $n=0$ and $n=N$ of the stripe. The slope of dispersion curves determines the circulation direction (chirality) of edge states. (b) Energy spectrum (left panel) and PR of corresponding eigenvectors (right panel) in a square lattice $N \times N=50 \times 50$ with a flux $\varphi= \pi/2$. }
\end{figure}

\subsection{Topological insulator laser: rate-equation model}
The topological insulator laser, considered in \cite{r4,r5}, is obtained by inserting optical gain in the microrings at the perimeter $\mathcal{P}$ of the square lattice and optical loss in the inner microrings to facilitate oscillation in the edge supermodes. Outcoupling is assumed at the edge microring at site $(1,1)$, which is coupled for example to a bus waveguide; see Fig.2(b). The laser system is described by the non-Hermitian model [compare with Eq.(2)]
\begin{equation}
i\frac{d \mathbf{c}}{dt}= (\mathcal{H}^{(Herm)} +\mathcal{H}^{(dis)}) \mathbf{c}+ i {\mathcal {L}} \mathbf{c}.
\end{equation}   
In Eq.(3), $\mathcal{H}^{(dis)}$ is an additional Hermitian contribution to the  Hamiltonian $\mathcal{H}^{(Herm)} $ that describes disorder $\Delta_{n,m}$ of the resonance frequencies of the microrings
\begin{equation}
(\mathcal{H}^{(dis)} \mathbf{c} )_{n,m}=\Delta_{n,m} c_{n,m}
\end{equation}
whereas $i \mathcal{L}$ is the non-Hermitian term that describes gain and loss in the system. Optical gain is here modeled using a standard rate equation model for class-B semiconductor lasers, which accounts for the slow carrier population dynamics and linewidth enhancement factor $\alpha$ \cite{r8,r10,r12,r16,r18}. The non-Hermitian term is given by
\begin{equation}
(\mathcal{L} \mathbf{c})_{n,m}=-\gamma c_{n,m} 
\end{equation}
for $(n,m) \notin \mathcal{P}$ and
\begin{eqnarray}
(\mathcal{L} \mathbf{c})_{n,m} & = & \frac{1}{2} \left( -\frac{1}{\tau_p}+ \sigma(N_{n,m}-1) \right) (1+i \alpha)c_{n,m} \nonumber \\
& - & \gamma_{out}c_{1,1} \delta_{n,1}\delta_{m,1}
\end{eqnarray}
for $(n,m) \in \mathcal{P}$, where $\mathcal{P}$ indicates the perimeter of the lattice, $\gamma$ is the linear loss rate in the inner rings, $\tau_p$ is the photon lifetime in the outer rings when the semiconductor has reached transparency, $N_{n,m}$ is the carrier density in the $(n,m)$ ring, normalized to the transparency value, $\alpha$ is the linewidth enhancement factor, $\sigma$ is proportional to the differential gain of the semiconductor, and $\gamma_{out}$ is the outcoupling loss of the $(1,1)$ microring. The equation for the carrier density $N_{n,m}$ reads
\begin{equation}
\frac{dN_{n,m}}{dt}=R-\frac{N_{n,m}}{\tau_s}-\frac{2}{\tau_s}(N_{n,m}-1)|c_{n,m}|^2
\end{equation}
where $R$ is the normalized injection rate of carriers ($R=R_{tr}= 1/ \tau_s$ at transparency), and $\tau_s$ is the carrier lifetime. The injection current is assumed to be uniform in all the rings of the perimeter $\mathcal{P}$.  Note that in the linearized regime (laser below or close to threshold) the linearized gain provided by the pumping in the rings is given by $g=(\sigma/2)(R \tau_s-1)$. We note that the limiting case of instantanesou gain saturation (class-A laser), used in \cite{r4}, can be retrieved by assuming $\tau_s \ll \tau_p$, which justifies adiabatic elimination of carrier density.

 \begin{figure}
\includegraphics[width=8.6cm]{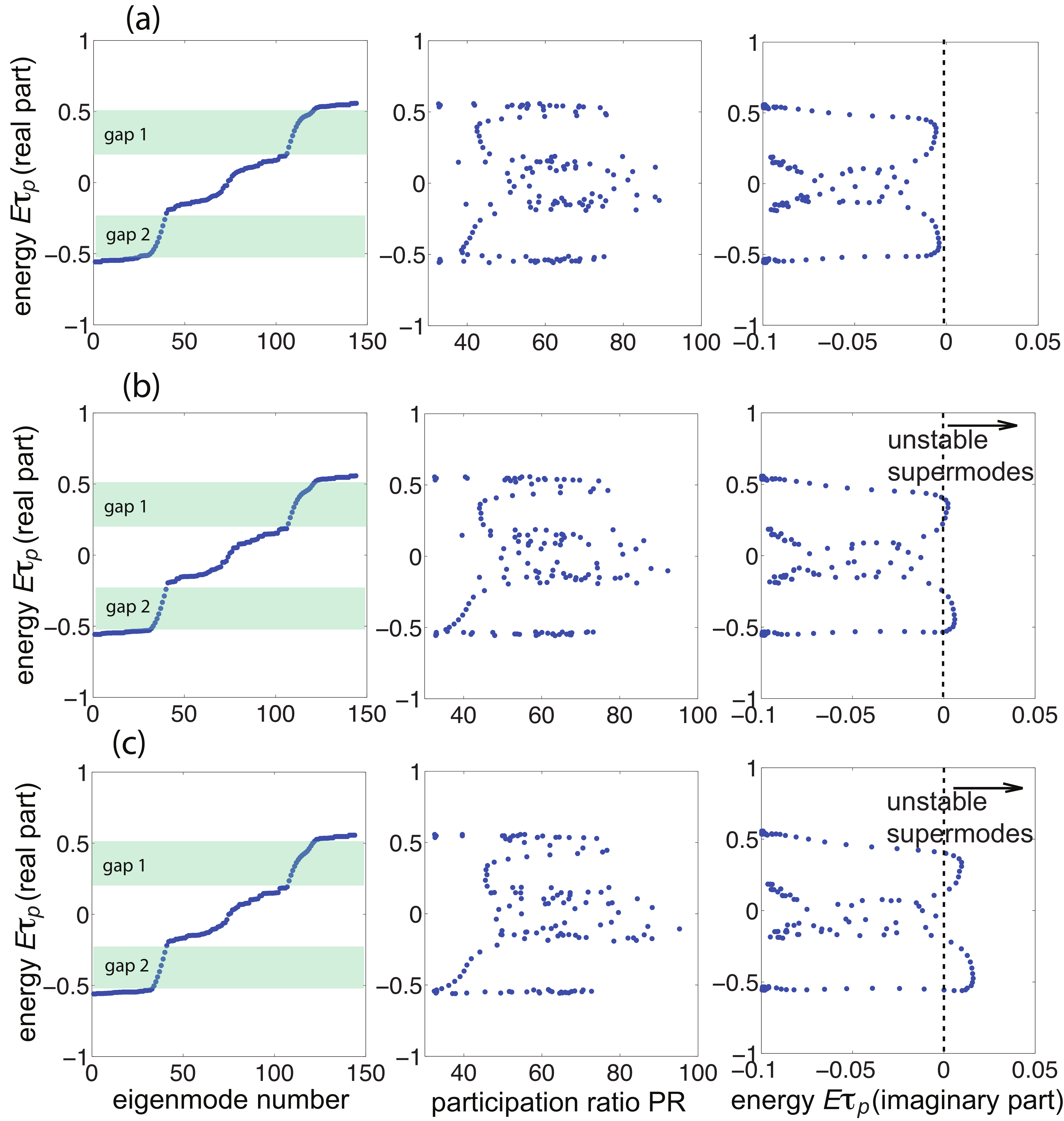}
\caption{ Numerically-computed eigenenergies (real and imaginary parts) and participation ratio of supermodes for the topological insulator laser ($N \times N=12 \times 12$) with linear gain and loss for increasing values of the normalized pump parameter $p$. Parameter values are: $\varphi=\pi/2$, $\alpha=3$, $\tau_p=40$ ps, $\sigma=6 \times 10^{11} \; {\rm s}^{-1}$, $\kappa \tau_p=0.2$, $\gamma = \kappa/2 $, and $\gamma_{out}=\kappa$. In (a)  $p=0.01$, in (b) $p=0.02$ and in (c) $p=0.03$. Supermodes with positive imaginary part of energy are unstable modes. Estimated threshold value is $p_{th} \simeq 0.014$. The pump parameters in (a), (b) and (c) correspond to normalized pump rates $R \tau_s=1.0425$,  $R \tau_s=1.0433$ and $R \tau_s=1.0442$, respectively. Note that such values are close to the transparency value $R \tau_s=1$ of the semiconductor, and very close to the threshold value of the single microring (accounting for its finite photon lifetime) $R \tau_s=1.0417$.}
\end{figure}

 \begin{figure}
\includegraphics[width=8.6cm]{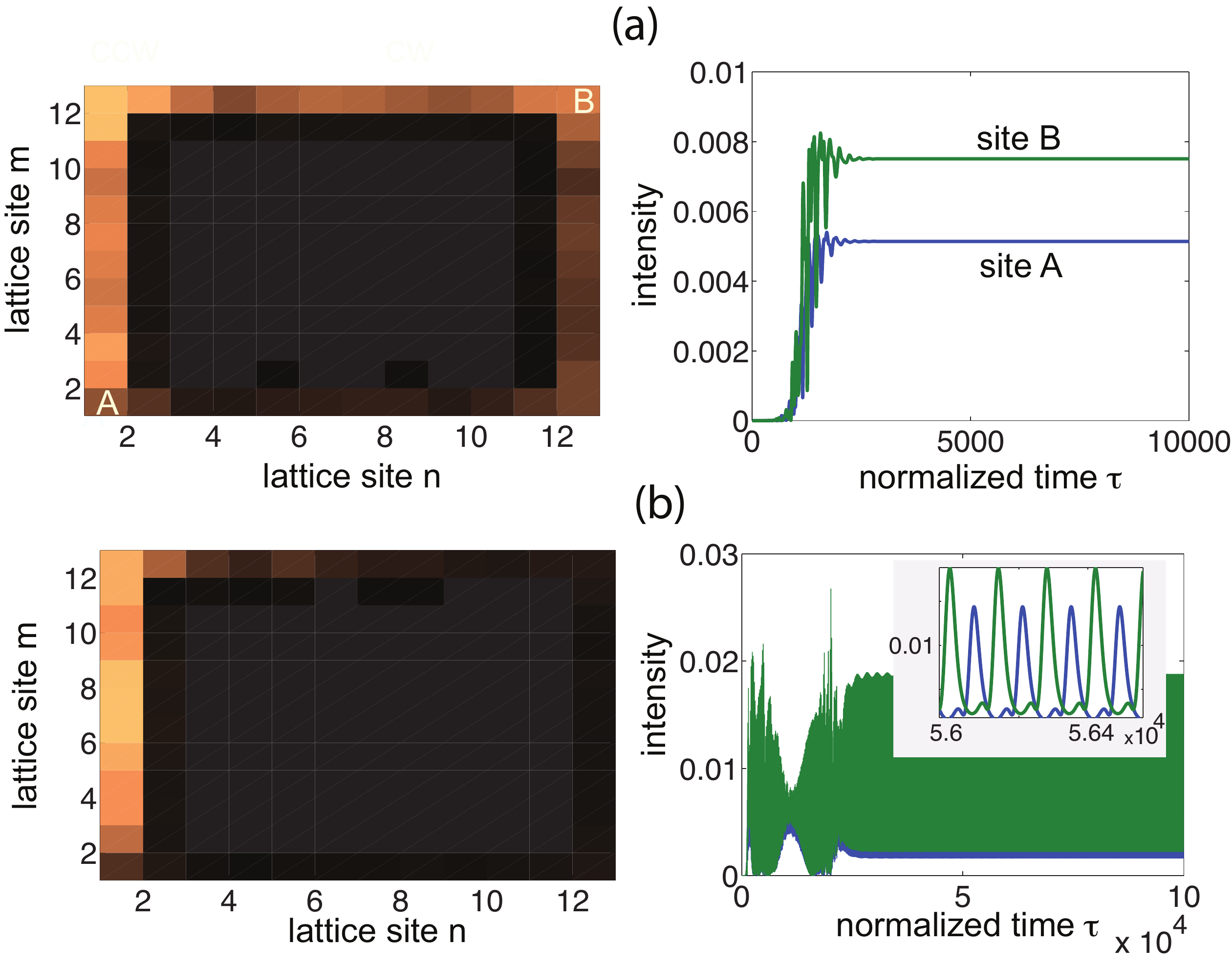}
\caption{ Transient switch-on dynamics of the topological insulator laser made of $12 \times 12$ microrings without disorder of resonance frequencies. Time is normalized to the cavity photon lifetime $\tau_p$ (i.e. $\tau=t / \tau_p$). Parameter values are: $\varphi=\pi/2$, $\alpha=3$, $\tau_p=40$ ps, $\sigma=6 \times 10^{11} \; {\rm s}^{-1}$, $p=0.02$, $\kappa \tau_p=0.2$, $\gamma = \kappa/2 $, $\gamma_{out}=\kappa$ and $\tau_s=8$ ps in (a) (virtual limit of class-A laser, $\tau_s / \tau_p=0.2$), $\tau_s=4$ ns in (b)  (class-B laser, $\tau_s / \tau_p=100$). The normalized pump parameter $p=0.02$ corresponds to a pump rate $R\tau_s=1.0433$, very close to its threshold value (see Fig.3). Left panels show the normalized intensity distribution $|c_{n,m}|^2$ in the lattice at final time [$t=10^4 \tau_p=0.4 \; \mu$s in (a) and $t=10^5 \tau_p=4 \; \mu$s in (b)]. Right panels show the temporal evolution of the field intensity at the vertex sites A (output coupler) and B.}
\end{figure}

 \begin{figure}
\includegraphics[width=8.6cm]{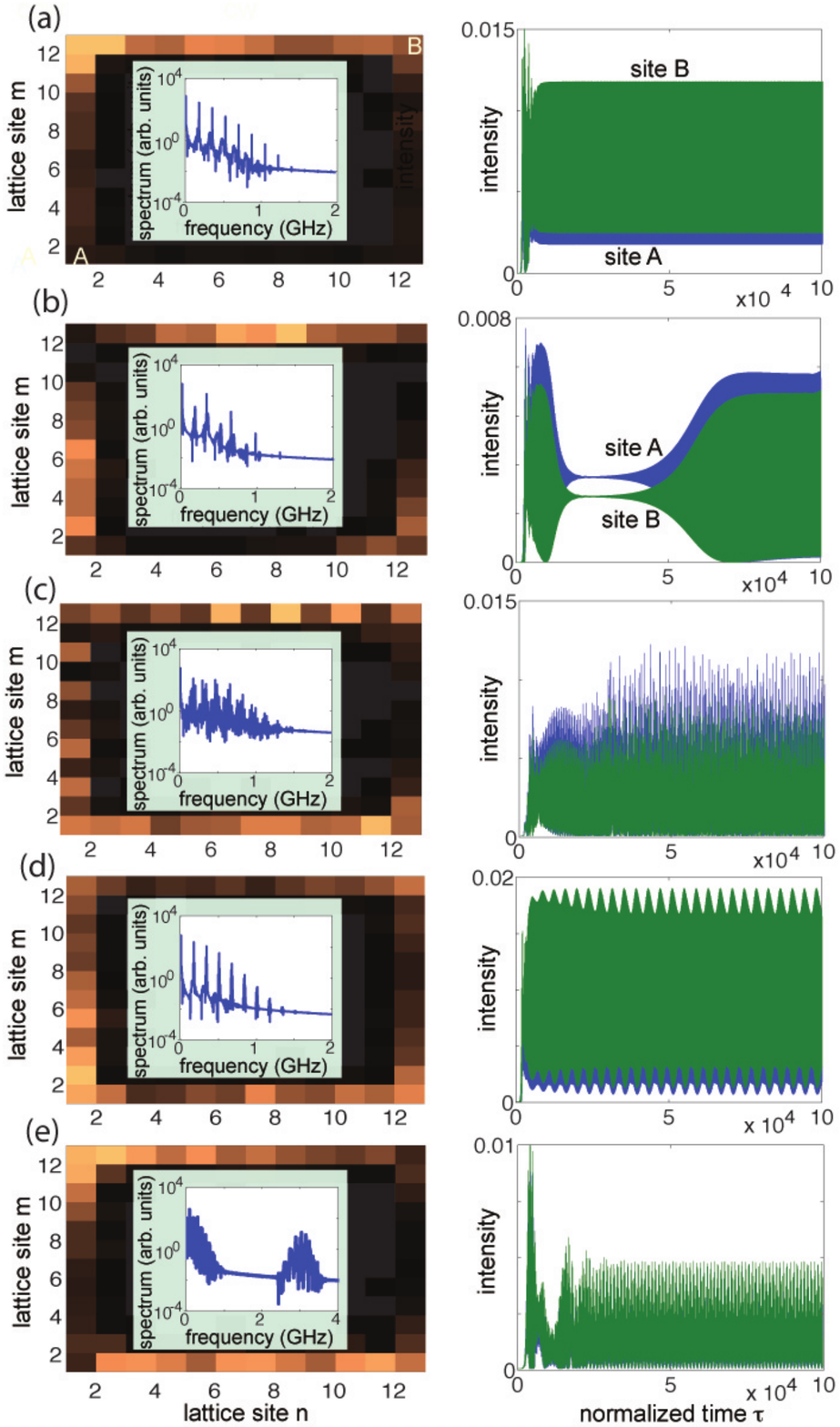}
\caption{ Same as Fig.4(b), but with disorder of resonance frequencies $\Delta_{n,m}$. The panels, form (a) to (e),  show different runs corresponding to different realizations of disorder, displaying distinct dynamical behavior. The insets in the left panels show the power spectrum of the temporal laser intensity at site A. In (e), the spectral peaks at higher frequencies (at around $\sim 3$ GHz) is ascribed to beating of counter propagating chiral edge modes. In 30 runs, steady-state oscillation after initial transient is observed in 4 runs (i.e. less than $20 \%$ times). The most probable outcome is the one in panel (a) (stable periodic limit cycle).}
\end{figure}

\subsection{Laser threshold of chiral edge supermodes}
The threshold condition of various edge supermodes of the lattice can be obtained by linear stability analysis of the non-lasing solution $c_{n,m}=0$ and $N_{n,m}=R \tau_s$ to Eqs.(3) and (7), which requires numerical computation of the eigenvalues of the $N \times N$ matrix, obtained from the Hermitian Hamiltonian $\mathcal{H}^{(Herm)}$ by including loss and linear gain. Typical examples of eigenvalues curves for increasing values of pumping are shown in Fig.3 in a lattice without disorder. Pumping is measured in terms of normalized excess pump parameter
\begin{equation}
p= (\sigma \tau_p/2)(R \tau_s-1)-1/2
\end{equation}
so that $p=0$ is the threshold value for a single pumped microring, decoupled from the others.
Unstable supermodes at the onset of lasing are those with a positive imaginary part of the eigenenergy. Note that some (albeit slight) discrimination in the threshold value of chiral edge supermodes is observed, with one supermode having the lowest threshold. Bulk modes, corresponding to larger PR,  have higher threshold value and thus are not lasing. Note that threshold condition does not depend on class-A or class-B dynamics. However, as discussed in the following, the above-threshold dynamics is strongly influenced by the ratio $\tau_s / \tau_p$, with a clear tendency to show dynamical instabilities in a realistic class-B laser model (even for pump rates very close to threshold).
\section{Topological insulator laser: dynamical instabilities}
Analytical form of stationary solutions to the laser equations, describing oscillation on a single edge chiral mode, is unfortunately unavailable, and hence it is not possible to determine  boundaries of stability of various laser supermodes like for simpler laser array models \cite{r8,r9,r10}. However, if a chiral edge supermode were a stable attractor of the dynamics (as claimed in \cite{r4}), we expect that, after initial relaxation oscillation  (transient laser switch on) stable emission should be observed, without self-pulsing or irregular temporal behaviors.\\ 
We numerically integrated the semiconductor laser rate equations, starting from random noise of field amplitudes and carrier density established by the pump rate, for parameter values $\tau_s=4$ ns, $\sigma=6 \times 10^{11} \; {\rm s}^{-1}$, $\alpha=3$, $\tau_p=40$ ps, $\kappa \tau_p = 0.2$, and for a square lattice made of $12 \times 12$ microrings. Lrger values of coupling, up to $\kappa \tau_p=3$, have been also considered. Such values can be regarded as realistic ones for the setup used in \cite{r5} (see also \cite{r18}). We compared the results obtained by realistic value of $\tau_s / \tau_p \simeq 100$  with the one corresponding to the virtual fast carrier relaxation rate $\tau_s / \tau_p=0.2 $, i.e. to  instantaneous gain saturation to mimic class-A dynamics as in \cite{r4}. The main important result is that, while in such a limiting case  after transient switch on steady-state operation in an edge supermode is most likely observed (according to \cite{r4}), by increasing the ratio $\tau_s/ \tau_p$ to a more realistic value, e.g. $\tau_s / \tau_p=100$ or in any case above 1 (class-B laser dynamics), irregular laser dynamics is much more likely to be observed even close to threshold. Such an irregular behavior is clearly ascribed to self-sustained relaxation oscillations, supermode competition and complex dynamics typical of semiconductor laser arrays with slow-gain dynamics and large linewidth enhancement factor. This means that, while topological edge supermodes are robust against disorder owing to their chirality, preventing oscillation of clusters of microrings, they are not immune to dynamical instabilities. Typical results of numerical simulations corroborating such a statement are shown in Figs.4 and 5 for ordered (Fig.4) and disordered (Fig.5) lattices. In the figures, time is normalized to the photon lifetime $\tau_p$. Figure 4 clearly shows that, while in the limit of class-A laser steady-state oscillation is reached after transient switch on [Fig.4(a)], corresponding to laser phase locking as stable fixed point, oscillatory dynamics is observed for the realistic class-B model, where a stable limit cycle is most likely observed [Fig.4(b)]. This behavior, i.e. transition from steady-state stable oscillation to oscillatory or irregular laser operation from class-A to class-B model,  is observed in most of numerical runs, starting from small random noise. The temporal behavior after transient laser switch on can change rather generally from run to run, especially in presence of disorder (depending on the specific realization of disorder $\Delta_{n,m}$). This is shown in Fig.5, where some examples of transient laser switch on dynamics, leading to a variety of attractors, are depicted. In the numerical simulations frequency detuning $\Delta_{n,m}$ is taken from a Gaussian distribution with zero mean and $0.28 \kappa$ variance. Rather generally, disorder seems to slightly combat the tendency of the ordered lattice to show oscillatory or irregular behavior, however a steady-state oscillation is observed in a minority of cases ( $<20\%$ probability in 30 runs). The most typical temporal trace corresponds to a stable periodic limit cycle [Fig.5(a)], however different behaviors can be observed, such as unstable fixed points [Fig.5(b)], chaotic states [Fig.5(c)], and quasi-periodic limit cycles [Fig.5(d)]. The spectral signatures of these oscillations are usually sidebands of the relaxation oscillation frequency \footnote{{For a single microring laser with photon lifetime $\tau_p$, the relaxation oscillation frequency is given by $\nu_{rel}\simeq (1 / 2 \pi) \sqrt{2p/ (\tau_s \tau_p)}$}.} (see the insets in Fig.5), which could be detected in optical power spectra measurements \cite{r11,rcazz1,rcazz2}. In some cases, higher-frequency irregular oscillations can be observed as well [Fig.5(e)], which arise from competition of counter-propagating chiral edge modes in the two gaps of Fig.3.  
Increasing the coupling constant $\kappa$ and losses $\gamma$ in the inner microrings scarcely help preventing dynamical instabilities, and irregular behavior of time traces is often observed. Other set of simulations have been performed, for example, by increasing $\kappa \tau_p$ from 0.2 up to 3, and the pump parameter $p$  up to $p=0.4$ (to reach laser threshold): even for such a stronger coupling, phase locking regime is often prevented by dynamical instabilities. The fact that the instabilities were not observed in the experiment of Ref.\cite{r5} is probably ascribed to the pulsed optical pumping regime: the pump pulses in \cite{r5} have a duration of $ \sim 10$ ns, which is comparable or even smaller than the characteristic time of relaxation oscillation period [see inset of Fig.4(b)].\par
It is worth comparing the topological insulator laser based on Hamiltonian (1), admitting chiral edge modes, with other methods of phase locking of laser arrays. Let us consider, for example, a linear chain of $N_d=4(N-1)$ pumped microrings placed along the perimeter $\mathcal{P}$ of a square lattice, see Fig.6(a). The geometry is similar to the topological insulator laser of Fig.1(b) but without the inner (lossy) rings.  Following  a recent proposal \cite{r19}, we assume nearest-neighbor coupling of microrings with coupling rates $\kappa_1$ and $\kappa_2$ for clockwise and counter-clockwise directions, with $\kappa_1  \geq \kappa_2$ . $\kappa_1=\kappa_2$ corresponds to ordinary Hermitian coupling and is somehow analogous to the topological insulator laser in the trivial topological phase $\varphi=0$. On the other hand,  asymmetric mode coupling $\kappa_1 > \kappa_2$  realizes a kind of non-Hermitian chiral transport along the perimeter $\mathcal{P}$ of the chain, which is robust against disorder owing to non-Hermitian Anderson delocalization transition \cite{r20,r21,r22,r23}. The limit $\kappa_2=0$ corresponds to unidirectional coupling of lasers in the array. Asymmetric coupling can be thought as arising from  a synthetic {\it imaginary} gauge field $h$, i.e. $ \kappa_{1,2}=\kappa \exp( \pm h)$) \cite{r20,r24}, rather than a real gauge field as in the topological insulator laser. An imaginary gauge field can be realized using antiresonance link rings that introduce unbalanced losses in their half perimeter \cite{r21,r25,r26}, rather than unbalanced phases. It is worth comparing the tendency to show dynamical instabilities in the topological insulator laser model of Refs.\cite{r4,r5} and in the laser array with non-Hermitian chiral modes based on asymmetric mode coupling. The rate equations for the latter laser array read
\begin{eqnarray}
\frac{dc_{\beta}}{dt} & = & \frac{1}{2} \left( -\frac{1}{\tau_p}+ \sigma(N_{\beta}-1) \right) (1+i \alpha)c_{\beta}-i \kappa_1c_{\beta-1} \nonumber \\
& - & i \kappa_2 c_{\beta+1} -(\kappa_1-\kappa_2+\gamma_{out} \delta_{\beta,1}) c_{\beta} \\
\frac{dN_{\beta}}{dt} & = & R-\frac{N_{\beta}}{\tau_s}-\frac{2}{\tau_s}(N_{\beta}-1)|c_{\beta}|^2
\end{eqnarray}
where $\beta=1,2,...,N_d$ and periodic (cyclic) boundary conditions $c_{\beta+Nd}(t)=c_{\beta}(t)$ apply. Outcoupling is at the microring $\beta=1$. In \cite{r19} analytical and numerical results showed that asymmetric mode coupling can fight the tendency of laser arrays to display dynamical instabilities as compared to ordinary symmetric (Hermitian) mode coupling. Figure 6 shows typical numerical results obtained for symmetric coupling ($\kappa_1=\kappa_2=\kappa$) and asymmetric coupling ($\kappa_1=\kappa$ and $\kappa_2<\kappa$) and for the same parameter values of $\alpha$, $\kappa$, $\tau_p$, $\gamma_{out}$, $p$ used in Fig.5 for the topological insulator laser. For ordinary Hermitian coupling $\kappa_1=\kappa_2=\kappa$, irregular oscillations are observed [Fig.6(b)]. This setting is similar to the topological insulator laser with trivial phase $\varphi=0$. However, for asymmetric mode coupling [Figs.6(c-d)] steady-state stable oscillation is observed. In 30 runs, corresponding to different realizations of disorder, we always observed steady-state oscillation, while for the topological insulator laser with the same parameter values (Fig.5) steady-state oscillation is observed in less than $20 \%$ cases. This result is mainly ascribable to the fact that non-Hermitian (asymmetric) mode coupling, besides of realizing chiral transport along the perimeter $\mathcal{P}$ which is robust against disorder, provides a better discrimination of array supermodes, favoring stable oscillation of the supermode traveling along the ring with the highest speed \cite{r19}.\\

\begin{figure}
\includegraphics[width=8.3cm]{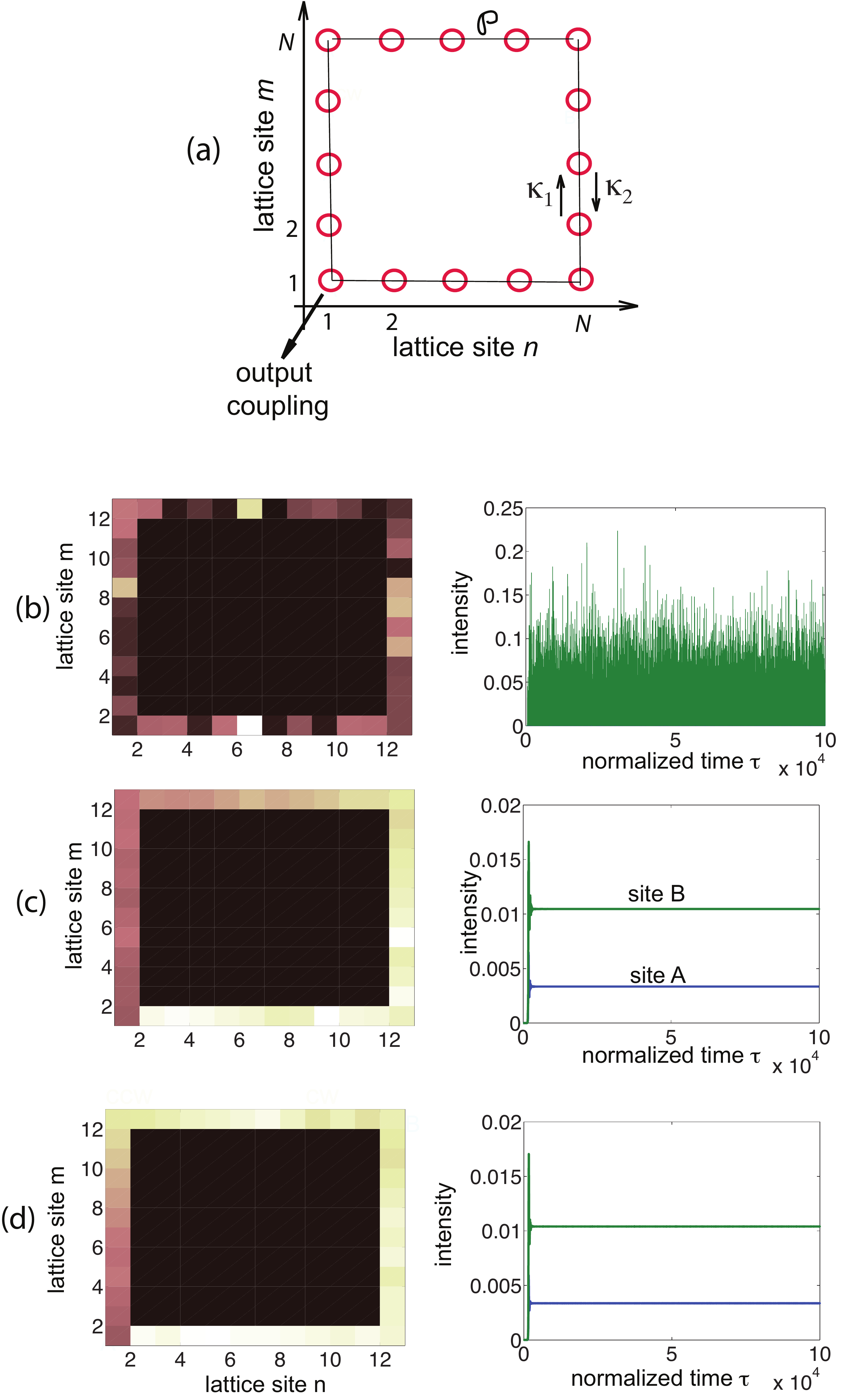}
\caption{ Laser phase locking based on asymmetric mode-coupling in a closed path $\mathcal{P}$. (a) Schematic of the laser chain. (b-d) Laser dynamics for $\kappa_1=0.2 / \tau_p$ and for (b) $\kappa_2= \kappa_1$ (Hermitian coupling), (c) $\kappa_2= 0.2 \kappa_1$, and (d) $\kappa_2=0$ (unidirectional coupling). Other parameters are as in Fig.5 (including disorder strength).}
\end{figure}

\section{Conclusions and discussion} 
Topological insulator semiconductor lasers, introduced and realized in recent works \cite{r4,r5}, have been proven to show robust transport of chiral edge modes, which are immune to disorder and imperfections in the lattice. A nontrivial topological phase, obtained by a synthetic magnetic flux based on anti resonant link rings \cite{r6,r7}, ensures higher laser slope efficiency and higher temporal coherence, preventing isolated (non-collective) oscillation of clusters observed  in the trivial topological phase. The theoretical model of the topological insulator laser, presented in \cite{r4}, assumed fast gain relaxation, which is valid for class-A lasers. In this regime the lowest threshold edge supermode stably oscillates above threshold and gain saturation suppresses other supermodes from lasing. However,  a minimal model of the topological insulator semiconductor laser should properly account for the slow relaxation rate of carrier density (class-B lasers) and for a non-negligible linewidth enhancement factor. Here we have shown that the complex field-carrier dynamics gives rise to dynamical instabilities, which are not captured in the simplified class-A laser model of \cite{r4}. Oscillatory instabilities are generally found for parameter values  that are of practical relevance. The spectral signatures of these intensity oscillations are side bands of the relaxation frequency, which should de observable in an experiment with continuous-wave (or quasi-continuous wave) pumping. Such a kind of instabilities are common in semiconductor laser arrays \cite{r8,r9,r10,r11,r12} and represent one of the main reasons that prevents stable oscillation of high-power laser arrays, at least without special cavity design \cite{r13,r14,r15}. Our results suggest that, while chiral edge lasing modes in topological insulator lasers are robust against disorder, they might not be immune to dynamical instabilities arising from complex carrier-field dynamics. Further experimental investigations could be devoted to characterize  the temporal behavior of topological insulator lasers. For future research, it would be of also major interest to combine the robustness against disorder of chiral edge transport with a laser design that could suppress or mitigate dynamical instabilities. A possibility could be to exploit {\it non-Hermitian} chiral transport, realized by a synthetic imaginary gauge field \cite{r20,r21,r22,r23}. Non-Hermitian mode coupling ensures chiral transport, robust against imperfections and disorder in view of the phenomenon of non-Hermitian Anderson delocalization \cite{r20,r23}, and can mitigate dynamical instabilities, as shown in a recent study \cite{r19}.  

\acknowledgments 
VK's research is partly supported by two ORAU grants entitled $^{\prime}$Taming Chimeras to Achieve the Superradiant Emitter$^{\prime}$ and $^{\prime}$Dissecting the Collective Dynamics of Arrays of Superconducting Circuits and Quantum Metamaterials$^{\prime}$, funded by Nazarbayev University. Funding from MES RK state-targeted program BR05236454 is also acknowledged.

\end{document}